\documentstyle[emulateapj,psfig]{article}

\newcommand{\be}{\begin{equation}}
\newcommand{\ee}{\end{equation}}
\newcommand{\ba}{\begin{eqnarray}}
\newcommand{\ea}{\end{eqnarray}}
\newcommand{\siml}{\lower4pt \hbox{$\buildrel < \over \sim$}}
\newcommand{\simg}{\lower4pt \hbox{$\buildrel > \over \sim$}}

\slugcomment{Accepted for publication in The Astrophysical Journal
Letters}

\begin{document}

\title{A model for the flaring radio emission in the double pulsar
system J0737-3039}

\author{Bing Zhang$^{1}$ and Abraham Loeb$^{2}$}

\affil{$^1$ Department of Physics, University of Nevada, Las Vegas, NV 89154 \\
$^2$ Astronomy Department, Harvard University, 60 Garden St.,
Cambridge, MA 02138}

\begin{abstract}
The binary J0737-3039{\bf A} \& {\bf B} includes two pulsars in a
highly relativistic orbit.  The pulsed radio flux from pulsar {\bf
B} brightens considerably during two portions of each orbit. This
phenomenon cannot be naturally triggered by the illumination of
$\gamma$-rays or X-rays from pulsar {\bf A} or the bow shock
around pulsar {\bf B}. Instead, we explain these periodic flares
quantitatively as episodes during which pairs from pulsar {\bf
A}'s wind flow into the open field line region of pulsar {\bf B}
and emit curvature radiation at radio frequencies within an
altitude of $\sim 10^8$ cm. The radio photons then travel through
{\bf B}'s magnetosphere and eventually reach the observer on the
other side of the pulsar. Our model requires that {\bf A}'s wind
be anisotropic and that {\bf B}'s spin axis be somewhat misaligned
relative to the orbital angular momentum. We estimate the expected
$\gamma$-ray and X-ray emission from the system.
\end{abstract}


\section{Introduction}

In the double pulsar system PSR J0737-3039{\bf A} \& {\bf B}
discovered recently (Burgay et al. 2004; Lyne et al. 2004), the
pulsed radio emission from {\bf B} brightens by orders of
magnitude near the orbital longitudes 210$^{\circ}$ and
280$^{\circ}$, while during most of the orbit, {\bf B}'s radio
emission is faint or non-detectable (Lyne et al. 2004;
Ramachandran et al. 2004). It has been heuristically suggested
that {\bf B}'s radio flares might be induced by the illumination
of {\bf A}, either in the form of particles, $\gamma$-rays, or
radio photons (Lyne et al. 2004; Jenet \& Ransom 2004). Assuming a
distance of $\sim 0.5$ kpc (Lyne et al. 2004), the system was
detected by {\em Chandra} to have an X-ray luminosity, $L_x \sim 2
\times 10^{30} ~{\rm erg~s^{-1}}$ (McLaughlin et al. 2004). It is
also within the EGRET point spread function of the unidentified
$\gamma$-ray source 3EG J0747-3412 (Hartmann et al. 1999), whose
luminosity is $\sim 2\times 10^{32} (\Delta \Omega_\gamma)~{\rm
erg ~s^{-1}}$ for the same distance and an unknown solid angle
$\Delta \Omega_\gamma$.

\section{Basic emission properties}

We first estimate the emission properties of both pulsars based on
their measured period $P$ and period derivative $\dot P$, by
ignoring the interaction between them.  Since the spindown power
of {\bf A} is much larger than that of {\bf B}, $\dot E_A \sim 5.9
\times 10^{33} ~{\rm erg~ s^{-1}} \gg \dot E_B \sim 1.6 \times
10^{30} ~{\rm erg~ s^{-1}}$, the interaction between the two
pulsars is expected to have a negligible effect on {\bf A}. The
emission of {\bf B} is, however, strongly influenced by the
interaction, as we show in \S3.

\noindent $\bullet$ {\bf Pulsar A}: The spin parameters $P_A=22.7$
ms, and $\dot P_A =1.74\times 10^{-18}$ (Burgay et al. 2004),
yield a polar cap surface magnetic field $B_{p,A} \sim 1.3\times
10^{10}$ G.  The empirical relation between $\gamma$-ray
luminosity $L_\gamma$ and $\dot E$ for known $\gamma$-ray pulsars
(i.e. $L_\gamma \propto \dot E^{1/2}$, Thompson 2003; Zhang \&
Harding 2000) yields $L_{\gamma,A} \sim 1.3 \times 10^{33} ~{\rm
erg~s^{-1}}$.  More adequate pulsar acceleration models suggest a
$\gamma$-ray luminosity in the range of $(0.3$--$1)\times 10^{33}
~{\rm erg~s^{-1}}$ (Zhang \& Harding 2000; Harding, Muslimov \&
Zhang 2002). This estimate is consistent with the observed
luminosity of the possibly related source 3EG J0747-3412, for
$\Delta \Omega_\gamma \sim (1 - 3)$. An associated X-ray
luminosity $L_{x,A} \siml 10^{-3} \dot E_A \sim 5 \times 10^{30}
~{\rm erg~s^{-1}}$ is expected from the full polar cap cascade
(Zhang \& Harding 2000), consistently with the empirical relation
discovered in other spin-powered pulsars (Becker \& Tr\"umper
1997) and the {\em Chandra} data (McLaughlin et al. 2004). Like
other millisecond pulsars, the radio emission of {\bf A} has a
wide double conal structure. Also the pulsar has a small
inclination angle between the spin and magnetic axes (Demorest et
al. 2004). The Goldreich-Julian (1969) pair flux from both
magnetic poles is $\dot N_{GJ,A} = 2.8 \times 10^{32} ~{\rm
s^{-1}} B_{p,10} P_{-2}^{-2} \sim 7 \times 10^{31} ~{\rm s}^{-1}$.

\noindent
$\bullet$ {\bf Pulsar B}: The spin parameters are $P_B=2.77$ s, and $\dot
P_B =0.88\times 10^{-15}$ (Lyne et al. 2004), yielding a polar cap surface
magnetic field of $B_{p,B} \sim 3.2\times 10^{12}$ G.  The pulsar polar gap
is likely controlled by the IC process, and the $\gamma$-ray luminosity in
this regime is $L_{\gamma,B} \sim 0.05 \dot E_B \sim 8\times 10^{28} ~{\rm
erg~s^{-1}}$ (Harding et al. 2002). The X-ray luminosity may be estimated
as $\sim 10^{-3} \dot E_B \sim 10^{27} ~{\rm erg~s^{-1}}$. Both values are
much smaller than the corresponding values for {\bf A}. The
Goldreich-Julian flux from both poles is $\dot N_{GJ,B} \sim 1.1 \times
10^{30} ~{\rm s^{-1}}$, and the expected pair multiplicity is $\kappa_B
\sim 0.1 \kappa_{B,-1}$ (Hibschman \& Arons 2001). The intrinsic pair
injection rate in the pulsar B magnetosphere is therefore
\begin{equation}
\dot N_{\pm} (B) \sim 10^{29} ~{\rm s^{-1}}.
\label{NpairB}
\end{equation}

\section{Physical mechanisms for {\bf B}'s radio flares\label{sec:int}}

Pulsar radio emission is broadly attributed to coherent emission by
electron-positron pairs in the pulsar magnetosphere. Our working hypothesis
is that during {\bf B}'s radio flares, there is a significant increase of
the pair injection rate into the radio emission region.

\subsection{$\gamma$-ray and X-ray precipitations}

We consider it unlikely that {\bf B} brightens because it is illuminated by
{\bf A}'s radio beam, since radio waves do not lead to production of new
e$^{+}$e$^{-}$ pairs.  Jenet \& Ransom (2004) suggested that the
$\gamma$-ray beam may coincide with the radio beam of {\bf A} and the
energetic $\gamma$-ray photons would trigger a pair cascade in {\bf B}'s
magnetosphere and lead to the radio flares. Putting aside the drawback that
the $\gamma$-ray and radio beams are usually misaligned in known
$\gamma$-ray pulsars (Thompson 2003), we demonstrate below that the
physical parameters of this scenario are implausible.

To maximize its effect, we assume that {\bf A}'s $\gamma$-ray
luminosity is close to its highest expected value, $L_{\gamma,A}
\sim 10^{33} ~{\rm erg~s^{-1}}$ beaming into a solid angle $\Delta
\Omega_\gamma \sim 1$. We also assume that the $\gamma$-ray
spectrum resembles that of known $\gamma$-ray pulsars, i.e. the
energy flux $\nu F_\nu$ is flat between 1 MeV and $\sim$ 30 GeV
(Thompson 2003). The range is bounded from above by the highest
photon energy capable of escaping from the polar cap cascade of
{\bf A} (see Eq. 29 in Zhang \& Harding 2000). We may then express
the injection spectrum as $E^2 \dot N(E) = 10^{33} / \log (3\times
10^4) \sim 10^{32} ~{\rm erg~s^{-1}}$. In order to produce a pair
in {\bf B}'s magnetosphere, the energy of a $\gamma$-ray photon,
$E_\gamma$, needs to satisfy $(E_\gamma / 2 m_e c^2) (B_\perp/B_q)
\geq \chi \sim 1/12$ (Ruderman \& Sutherland 1975), where $B_\perp
= B \sin \theta_{kB}$ and $\theta_{kB}$ is the angle between the
$\gamma$-ray momentum and the magnetic field ${\vec B}$. Again, to
maximize the effect we adopt $\sin \theta_{kB} \sim 1$. For a pure
dipole field, $B \sim B_p (R/r)^3$ (where $R$ is the stellar
radius and $r$ is the radius at which pairs are produced), the
threshold $\gamma$-ray energy is $E_{\gamma,th} \sim 1.2 ~{\rm
MeV} (r/R)^3$.  Given $E_{\gamma,max}$, the maximum radius for
pair production is then $r_{max}/R = (3\times 10^4/1.2)^{1/3} \sim
29$. The external pair cascade process is dominated by the
synchrotron radiation of the higher generation pairs. Since the
photon energy decreases by a factor 16 in each generation, each
primary $\gamma$-ray with energy $E_\gamma$ can generate
$2^{\zeta}$ pairs, where $\zeta = [\log(E_{\gamma} /
E_{\gamma,th}) / {\log (16)}]+1$ (e.g. Zhang \& Harding 2000). The
radius-dependent solid angle (including both poles) of the open
field line region is $\Delta \Omega_{open} (r) = 4\pi^2 r/cP$, and
so the pair injection rate into {\bf B}'s open field region due to
$\gamma$-rays from {\bf A} is
\begin{eqnarray}
\dot N_\pm (A\rightarrow B) & = & \int_R^{r_{max}}
\left[\int_{E_{\gamma,th}(r)}^{E_{\gamma,max}}
2^\zeta \dot N(E) dE \right] \nonumber \\
& \times & \frac{\Delta\Omega_{open}(r) r dr}{\Delta\Omega_\gamma
d_{AB}^2} ~~ \sim ~ 10^{26} ~{\rm s^{-1}}~,
\label{Npair1}
\end{eqnarray}
where the distance between the two pulsars is $d_{AB} \sim 9
\times 10^{10}$ cm (Lyne et al. 2004). This rate is negligible
compared with the intrinsic pair injection rate of {\bf B} itself
(Eq. \ref{NpairB}). The total pair injection rate into {\bf B}'s
magnetosphere (including the closed field line region) is $2\times
10^{29} ~{\rm s^{-1}}$ (calculated by replacing
$\Delta\Omega_{open}(r)$ by $4\pi$ in Eq. \ref{Npair1}), but it is
believed that pairs in the closed field line region can not
contribute to the observed coherent emission from
pulsars\footnote{When estimating the pair injection rates (Eqs.
\ref{NpairB}, \ref{Npair1}), a pure dipolar geometry was assumed
for {\bf B}. In reality, {\bf B}'s magnetosphere is severly
distorted by {\bf A}'s wind (Arons et al. 2004; Lyutikov 2004),
but our conclusion remains unchanged when the distortion effect is
included.}.

{\bf A}'s wind is terminated by the magnetic stress within {\bf
B}'s magnetosphere through a bow shock. The distance of the bow
shock from {\bf B} is $d_{sB}=({8 \mu_B^2 d_{AB}^2 c}/{\dot
E_A})^{1/6} \sim 5\times 10^9 ~{\rm cm}$ (Arons et al. 2004),
where $\mu_B \sim 3.75 \times 10^{29}$ cgs is the magnetic moment
of {\bf B}. The bow shock produces a $\gamma$-ray luminosity of
$\sim 10^{30}~{\rm erg~s^{-1}}$ at $\sim 20$ MeV (Granot \&
M\'esz\'aros 2004). The ratio between the $\gamma$-ray flux from
{\bf A} and this component is $\sim (10^{33}/10^{30})\times
(d_{sB}/d_{AB})^2 \sim 3$, and so this component does not increase
significantly the pair abundance in {\bf B}'s magnetosphere. The
bow shock also produces X-rays, which may interact with the high
energy $\gamma$-rays from the polar cap to produce pairs (see e.g.
Zhang 2001; Harding et al. 2002). However, for the estimated X-ray
luminosity $10^{29}~{\rm erg~s^{-1}}$ (Granot \& M\'esz\'aros
2004), the number density of X-ray photons in the open field line
region is only $\sim 10^9 ~{\rm cm^{-3}}~(\epsilon_x / 0.1~{\rm
keV})^{-1}$, leading to an optical depth $\tau_{\gamma\gamma} \sim
10^{-5}$ for a $\gamma$-ray traveling through the entire
magnetosphere.  The X-rays coming directly from {\bf A} have an
effect that is smaller by a factor $\sim (10^{30}/10^{29})\times
(d_{sB}/d_{AB})^2 \sim 0.03$.

We therefore conclude that the radio flares from {\bf B} are not
due to $\gamma$-ray or X-ray precipitations from either pulsar
{\bf A} or from the bow shock around pulsar {\bf B}.

\subsection{Leakage of {\bf A}'s wind into {\bf B}'s magnetosphere}

The luminous {\bf A} wind distorts {\bf B}'s magnetosphere into a shape
analogous to the Earth magnetosphere as it is combed by the solar
wind. According to the numerical simulation of Arons et al. (2004), the
vertical radius of the magnetospheric sheath is $l \sim 7.5 \times 10^9$
cm.  Interpreting {\bf A}'s eclipse as synchrotron self-obsorption in the
shocked {\bf A} wind requires that the pair multiplicity of {\bf A} be as
high as $\kappa_{A} \sim 10^6$ (Arons et al. 2004; Lyutikov 2004). This
multiplicity value is much larger than in standard cascade theory, $\la
100$ (Hibschman \& Arons 2001). Hereafter, we normalize $\kappa_A$ by
$10^6$, although our principal conclusions remain valid at lower values.
The total number of the pairs deposited to the bow shock region from {\bf
A}'s wind is
\begin{eqnarray}
\dot N_{\pm} (A \rightarrow sh) & = & \dot N_{GJ,A}
\kappa_{A} \frac{\pi l^2}{\Delta\Omega_{w,A} (d_{AB}-d_{sB})^2} \nonumber
\\
& \sim & 1.2\times 10^{35} ~{\rm s}^{-1} ~\kappa_{A,6}
(\Delta\Omega_{w,A}/4\pi)^{-1},
\label{Npair2}
\end{eqnarray}
where $\Delta\Omega_{w,A}$ is the unknown solid angle of {\bf A}'s wind.

It has been suggested that {\bf B} spin axis aligns in the
direction almost perpendicular to the orbital plane due to the
external torque exerted by {\bf A}'s wind (Demorest et al. 2004;
Arons et al. 2004). Since the line of sight (which sweeps across
{\bf B}'s radio beam) is offset by $3^{\circ}$ from the orbital
plane (Lyne et al. 2004; Kaspi et al. 2004), {\bf B}'s magnetic
axis must be oriented at a small angle relative to the orbital
plane.  It is therefore likely that {\bf A}'s wind would directly
stream into at least one of the open field regions of {\bf B}. The
``leakage'' may be realized through resistive effects (see
Lyutikov 2004 and references therein). The analogy to the solar
wind interaction with the Earth's magnetosphere suggests that the
pairs gain access near {\bf B}'s magnetic pole. When this happens,
some fraction of the pairs in {\bf A}'s wind can directly slide
into {\bf B}'s magnetosphere. At the same time the open field
region in the ``day'' side facing {\bf A} is greatly broadened due
to the ram pressure of the wind. In principle, the stream from
{\bf A}'s wind may encounter {\bf B}'s pair stream (``wind'')
along its path. The distance from {\bf B} where the pressures of
the two streams balance, $d_{bB}$, can be found by equating
$\eta_B \dot E_{B}/(c \Delta \Omega_{w,B} d_{bB}^2) = \dot E_{A} /
[c \Delta \Omega_{w,A} (d_{AB}-d_{dB})^2]$, where $\Delta
\Omega_{w,B} \propto d_{bB}$ is the solid angle of {\bf B}'s wind,
and {\bf B}'s wind luminosity is smaller than the spindown power
by a factor of $\eta_B<1$ due to pair screening. We get $d_{bB}
\sim 1.5\times 10^9 ~{\rm cm}~ \eta_B^{1/2} (\Delta \Omega_{w,A} /
\Delta \Omega_{w,B})^{1/2} < d_{sB}$. Once pairs from {\bf A}'s
bow shock leak into {\bf B}'s magnetosphere, they will stream all
the way down to {\bf B}'s surface due to the external pressure
gradient from above.

We parameterize the pair injection rate into {\bf B}'s
open field line region as
\begin{equation}
\dot N_{\pm} (sh \rightarrow B)
\sim 1.2\times 10^{34} ~{\rm s}^{-1} ~\eta_{-1} \kappa_{A,6}
(\Delta\Omega_{w,A}/4 \pi)^{-1},
\label{Npair3}
\end{equation}
where $\eta = 0.1 \eta_{-1}$ is the fraction of all pairs from
{\bf A}'s bow shock that enter this region.  The resulting
injection rate is orders of magnitude larger than the intrinsic
pair injection rate from {\bf B} in equation (\ref{NpairB}), even
if $\kappa_{A,6}$ is smaller than unity. We therefore suggest that
{\it these pairs are the catalyst for {\bf B}'s radio flares}. A
natural coherent mechanism would be the two-stream instability
between the downstream {\bf A} wind and the upstream {\bf B} wind.
The radio photons emitted by these pairs travel downwards, but
owing to the curved dipole geometry of the magnetic field that is
responsible for their production, they are not blocked by the
neutron star. Rather, the radio waves pass through the inner
magnetosphere and the gravitational field of {\bf B} and
eventually reach the observer on the other side of the neutron
star.

The observed double conal emission of {\bf B} is separated by
about 0.03 s (Lyne et al. 2004). Since the open field region is
nearly oriented towards the observer, the radio beam opening angle
is $\theta_{b} \sim 2\pi(0.03/2.77) \sim 0.07$. Based on the
dipole field geometry and the gravitational bending effect, this
angle should be $\theta_b=2(\theta_t-\theta_g)$, where
$\theta_t=(3/2) \theta_r = (3/2) (r/r_L)^{1/2}$ is the angle
between the tangent of the field line at the emission point and
the magnetic polar direction, and $\theta_g=2 r_{\rm Sch}/b$ is
the gravitational deflection angle. Here $r_{\rm Sch}=2 G M/c^2
\sim 3.7\times 10^5$ cm is the Schwarzschild radius of the neutron
star for a mass of 1.25 $M_\odot$ (Lyne et al. 2004), and $b \sim
r(\theta_t-\theta_r)$ is the photon's impact parameter from the
star. Due to {\bf A}'s wind pressure, the field lines on the
``daytime'' side are broadened. We still assume a dipolar
geometry, but take the light cylinder radius $r_L$ to be
comparable to the vertical size of the magnetic sheath, $r_L \sim
l \sim 7.5 \times 10^9$ cm. This gives $\theta_r \sim 0.12
r_8^{1/2}$, $\theta_t \sim 0.17 r_8^{1/2}$ and $\theta_g \sim 0.13
r_8^{-3/2}$, where $r_8=r/10^8$ cm. We therefore obtain $\theta_b
\sim 2(0.17 r_8^{1/2} - 0.13 r_8^{-3/2})$, or $\theta_b \sim 0.08$
for $r_8 \sim 1$. Thus $r_8 \sim 1$ is required from the data for
such a geometry.

The average Lorentz factor of the pairs that enter {\bf B}'s
magnetosphere could be estimated from the relation $\gamma_\pm m_e
c^2 \dot N_{GJ,A} \kappa_A = \dot E_A$. In pulsar magnetospheres,
the synchrotron cooling time scale for pairs is very short so that
the perpendicular component of the pair energy is lost
instantaneously. The parallel Lorentz factor of the pairs is
$\gamma_{\pm,\parallel} = \xi_\parallel \gamma_{\pm} \sim 210
\kappa_{A,6}^{-1} \xi_\parallel$, where $\xi_\parallel\leq 1$ is a
geometric factor depending on the incidence angle of the pairs
relative to the magnetic field (Zhang \& Harding 2000). At a
height $r \sim 10^8 r_8$ cm, the curvature radius of the field
line for $r_L \sim l$ is $\rho \sim 1.2\times 10^9 r_8^{1/2}$ cm,
and the typical curvature radiation frequency of the pairs is
\begin{equation}
\omega_c =\frac{3}{2} \frac{\gamma_{\pm,\parallel}^3c}{\rho} \sim
350 ~{\rm MHz} ~  \kappa_{A,6}^{-3} \xi^3_\parallel r_8^{-1/2}~.
\label{omegac}
\end{equation}
The characteristic plasma frequency for the two stream instability
is $\omega_p=\gamma_{\pm,\parallel}(4 \pi n' e^2/m_e)^{1/2}$
(Ruderman \& Sutherland 1975; Medvedev \& Loeb 1999), where $n'
\sim \dot N_\pm(sh \rightarrow B)/\gamma_{\pm,\parallel}\pi r^2
\theta_r^2 c$ is the co-moving down-stream plasma density. If
$\omega_p < \omega_c$ (requiring $\kappa_{A,6} < 0.25
\xi^{5/6}_\parallel r_8^{1/3}$), then the two-stream instability
greatly amplifies the curvature radiation (Ruderman \& Sutherland
1975). Given the uncertainties in the value of $\kappa_{A,6}$ and
$\xi_\parallel$, the typical enhanced curvature radiation
frequency is consistent with the 680-3030 MHz band at which {\bf
B} is detected (Lyne et al. 2004). We therefore conclude that the
emission altitude is $\sim 10^8$ cm, a value consistent with the
emission altitudes derived for other pulsars (Kijak \& Gil 2003).
We note that although some models of pulsar radio emission (e.g.
Lyutikov, Blandford \& Machabeli 1999) disfavor emission from
regions near the magnetic pole, the observed narrow beam calls for
this geometry, and we speculate that the two-stream instability
discussed above plays an essential role in achieving the strong
coherence of the emission.

\subsection{Consequences of the model}

There are several direct consequences of our model.  First, the pairs
eventually deposit energy onto {\bf B}'s polar cap region, making {\bf B} a
strong X-ray emitter as well. The pairs lose energy via curvature radiation
and IC as they stream towards {\bf B}'s surface, and both braking processes
become more significant near the star surface. The curvature radiation
power $\dot \gamma_{CR} m_e c^2 = -(2c/3)(e^2/\rho) \gamma_\pm^4 \sim
10^{-7} ~{\rm erg~s^{-1}} r_6^{-1/2}$, is negligible. The IC braking is,
however, important. With $\gamma_\pm \sim 210$, the IC process is not
initially in the resonant regime, i.e. $\gamma_\pm \epsilon_x > (B_p/B_q)
m_e c^2 \sim 35~{\rm keV} (B_p/3\times 10^{12} ~{\rm G})$, where
$\epsilon_x$ is the typical X-ray photon energy (Eq. \ref{epsx}), and $B_q
= 4.414 \times 10^{13}$ G is the critical magnetic field. However, even in
the non-resonant regime, the IC power $\dot \gamma_{CR} m_e c^2 = -(4/3)
\gamma_\pm^2 c \sigma_{T} U_{ph}$ is large for typical parameter values, so
that the pairs undergo significant deceleration at $r<3R$. The IC process
then enters the resonant regime, and the braking power becomes even
larger. As the pairs hit {\bf B}'s surface, their typical Lorentz factor is
of order a few. However, the IC $\gamma$-rays are beamed towards the
surface, so that their energy is deposited on the near polar cap region as
well. We therefore assume that all the initial pair energy is deposited on
the polar cap region and later radiated as X-rays. The X-ray luminosity can
then be estimated as
\begin{eqnarray}
L_{x,B} & \sim & \dot N_{\pm} (sh \rightarrow B)
\gamma_{\pm,\parallel} m_e c^2
\nonumber \\
& \sim &
2.1\times 10^{30} ~{\rm erg~s^{-1}}~ \eta_{-1} \xi_\parallel
(\Delta\Omega_{w,A}/4\pi)^{-1}~,
\label{LxB}
\end{eqnarray}
yielding a value that is consistent with the {\em Chandra} observations
(McLaughlin et al. 2004) for $\eta \sim 0.1$ during the phases
when {\bf B} is illuminated by {\bf A}'s wind. The typical temperature of
the polar cap is $T_{pc} \sim (Ll/ \sigma \pi R^3)^{1/4} \sim 3.1\times
10^6 ~{\rm K} \eta_{-1}^{1/4} \xi_\parallel^{1/4}
(\Delta\Omega_{w,A}/4\pi)^{-1/4}$, with a typical photon energy
\begin{equation}
\epsilon_x \sim 2.8 k T_{pc} \sim 0.7 ~{\rm
keV}~\eta_{-1}^{1/4}\xi_\parallel^{1/4} (\Delta\Omega_{w,A}/4\pi)^{-1/4} ~.
\label{epsx}
\end{equation}
This X-ray component should be modulated at {\bf B}'s spin period of 2.77
s, but may not be modulated with orbital phase due to the long thermal
inertial time of neutron stars (Eichler \& Cheng 1989).

At the higher altitudes of $r\sim 10^7$cm, the IC photons produced by the
pairs will not be blocked by the neutron star and can reach the
observer. Their typical energy is $\epsilon_\gamma \sim \gamma_\pm^2
\epsilon_x \sim 30$ MeV. The number density of X-ray photons is $n_x \sim
L_{x,pc} / (\epsilon_x 2 \pi r^2 c) \sim 2\times 10^{11}~{\rm cm}^{-3}
\eta_{-1}^{3/4} r_7^{-2}$, giving a Thomson optical depth $\tau_{\rm IC} =
n_x r \sigma_T \sim 1.3\times 10^{-6} \eta_{-1}^{3/4} r_7^{-1}$. The
$\gamma$-ray luminosity is $L_{\gamma,B'} \sim 30~{\rm MeV} \times
\dot N (sh \rightarrow B) \times \tau_{\rm IC} \sim 7.5\times 10^{23}
~{\rm erg~s^{-1}}$, much lower than the intrinsic $\gamma$-ray
luminosity of {\bf B}, which itself is much lower than the
$\gamma$-ray luminosity of {\bf A}.

Since we have attributed {\bf B}'s radio flares to the interaction
between {\bf A}'s wind and the open field region of {\bf B}, the
geometric model of Jenet \& Ransom (2004) which is based on {\bf
A}'s radio beam illuminating {\bf B}, has to be modified (keeping
in mind that {\bf A}'s wind should be wider than its radio beam,
allowing for a temporal mismatch between the two as reported
[Ransom et al. 2004]). In order to obtain two specific orbital
phases during which {\bf B} flares, we require that {\bf A}'s wind
be anisotropic, and that {\bf B}'s spin axis be somewhat
misaligned relative to the orbital angular momentum. The lack of a
detection of inter-pulses supports the latter assumption. The
suggestion of two additional orbital episodes of enhanced weak
emission (Ramachandran et al. 2004) can be interpreted as the
interaction of {\bf A}'s wind with {\bf B}'s second magnetic pole
(which has the less favorable orientation). For the two bright
phases, the roughly-equal double-component radio profile near
longitude 280$^{\circ}$ (Lyne et al. 2004) is interpreted as {\bf
A}'s wind streaming directly into {\bf B}'s open field region,
while the weak precursor followed by an intense main pulse at
longitude 210$^{\rm \circ}$ (Lyne et al. 2004) can be explained if
{\bf A}'s wind only covers partially {\bf B}'s open field region.
For the other two dim phases (Ramachandran et al. 2004), either
the wind interaction region only covers a small fraction of the
open field region, or the line of sight only grazes the emission
region. A more refined geometric model could be developed from
additional radio data, including the consequences of geodetic
precession (Jenet \& Ransom 2004).

In our model, the radio waves travel through near-surface closed field
regions where the magnetic fields are stronger and the plasma is denser. At
closest approach to the neutron star, the local magnetic field is $\sim
1.6\times 10^{10}~{\rm G}$. The local plasma frequency is
$\omega_{p,c}=(4\pi n_{\pm,c} e^2/ \gamma_{\pm,c} m_e)^{1/2} \sim 110~{\rm
MHz}~ \kappa_{B,c}^{1/2} \gamma_{\pm,c,2}^{-1/2}$, where $n_{\pm,c}$,
$\kappa_{B,c}=n_{\pm,c}/n_{GJ}$, and $\gamma_{\pm,c}$ are the number
density, multiplicity and typical Lorentz factor of the pairs in the closed
field region (with these pairs being seeded by the $\gamma$-rays emanating
from {\bf A}, the bow shock, and the radio emission region). The plasma
cutoff frequency is much lower than the observed frequency, but could
introduce a large dispersion measure for {\bf B}.  The propagation through
the plasma may also lead to novel polarization signatures.

\section{Summary}

We have interpreted the periodical radio re-brightening of {\bf B}
as episodes during which pairs from {\bf A}'s wind flow into the
open field line region of {\bf B} and emit curvature radiation at
an altitude of $\sim 10^8$ cm. The pair multiplicity required to
explain {\bf A}'s eclipse as synchrotron absorption in the bow
shock around {\bf B}'s magnetosphere, yields the required radio
frequency for the curvature radiation in {\bf B}'s magnetoshpere
(Eq. \ref{omegac}). The radio photons travel through {\bf B}'s
magnetosphere and gravitational potential and eventually reach the
observer on the other side of the pulsar. Our model requires that
{\bf A}'s wind be anisotropic, and that {\bf B}'s spin axis be
somewhat misaligned with the orbital angular momentum.

The system has several components of high energy emission.  The
dominant $\gamma$-ray source is {\bf A}, whose luminosity
$L_{\gamma,A}\sim (0.3-1)\times 10^{33}~{\rm erg~s^{-1}}$ is
consistent with that of the unidentified EGRET source 3EG
J0747-3412. The source is expected to have a 22.7 ms period which
could be detected by the future {\em GLAST} mission
(http://glast.gsfc.nasa.gov/). In the X-ray band, there are three
sources that could account for the luminosity measured by the {\em
Chandra} satellite, $L_x\sim 2\times 10^{30} ~{\rm erg~s^{-1}}$.
They are: {\it (i)} polar cap heating and cascade emission from
{\bf A} (Zhang \& Harding 2000); {\it (ii)} {\bf B}'s polar cap
heating by {\bf A}'s wind (this work); and {\it (iii)} emission by
the interstellar medium shocked by {\bf A}'s wind (Granot \&
M\'esz\'aros 2004). The first two components should be pulsed with
the corresponding pulsar periods (22.7 ms and 2.77 s),
respectively, and should possess a thermal spectral component.
Further observations with longer exposure times are needed to
separate these components.

\acknowledgements

We thank Alice Harding, Scott Ransom and Peter M\'esz\'aros for useful
comments on the manuscript. A.L. thanks Vicky Kaspi for initiating his
interest in this problem.  This work was supported in part by NASA grants
NNG 04GD51G (for B.Z.), NAG 5-13292 (for A.L.), and by NSF grants
AST-0071019, AST-0204514 (for A.L.).


\begin{thebibliography}{}
\bibitem[]{} Arons, J., Backer, D. C., Spitkovsky, A. \& Kaspi, V.
2004, to appear in Proceedings of the 2004 Apsen Winter Conf. on
Astrophysics ``Binary Radio Pulsars'', ASP Conf Series, (eds.)
F. Rasio \& I. Stairs, (astro-ph/0404159)
\bibitem[]{} Becker, W. \& Tr\"umper, J. 1997, A\&A, 326, 682
\bibitem[]{} Burgay, M. et al. 2004, Nature, 426, 521
\bibitem[]{} Demorest, P. et al. 2004, ApJL, submitted
(astro-ph/0402025)
\bibitem[]{} Eichler, D. \& Cheng, A. F. 1989, ApJ, 336, 360
\bibitem[]{} Goldreich, P. \& Julian, W. H. 1969, ApJ, 157, 869
\bibitem[]{} Granot, J. \& M\'esz\'aros, P. 2004, ApJ, 609, L17
\bibitem[]{} Harding, A. K., Muslimov, A. G. \& Zhang, B. 2002, ApJ,
576, 366
\bibitem[]{} Hartman, R. C. et al. 1999, ApJS, 123, 79
\bibitem[]{} Hibschman, J. \& Arons, J. 2001, ApJ, 554, 624
\bibitem[]{} Jenet, F. A. \& Ransom, S. M. 2004, Nature, 428, 919
\bibitem[]{} Kaspi, V. M. et al. 2004, ApJL, submitted
(astro-ph/0401614)
\bibitem[]{} Kijak, J. \& Gil, J. 2003, A\&A, 397, 969
\bibitem[]{} Lyne, A. et al. 2004, Science, 303, 1153
\bibitem[]{} Lyutikov, M. 2004, MNRAS, submitted (astro-ph/0403076)
\bibitem[]{} Lyutikov, M., Blandford, R. D. \& Machabeli, G.
1999, MNRAS, 305, 338
\bibitem[]{} McLaughlin, M. A. et al. 2004, ApJ, 605, L41
\bibitem[]{} Medvedev, M. V. \& Loeb, A. 1999, ApJ, 526, 697
\bibitem[]{} Ramachandran, R., Backer, D. C., Demorest, P., Ransom,
S. M. \& Kaspi, V. M., 2004, ApJ, submitted (astro-ph/0404392)
\bibitem[]{} Ransom, S. M. et al. 2004, (astro-ph/0404341)
\bibitem[]{} Ruderman, M. \& Sutherland, P. G. 1975, ApJ, 196, 51
\bibitem[]{} Thompson, D. J. 2003, (astro-ph/0312272)
\bibitem[]{} Zhang, B. 2001, ApJ, 562, L59
\bibitem[]{} Zhang, B. \& Harding, A. K. 2000, ApJ, 532, 1150
\end{thebibliography}
\end{document}